\documentclass[sigconf]{acmart}

\usepackage{amsmath}
\usepackage{enumitem}
\usepackage{multirow}
\usepackage{colortbl}

\newcommand{\loss}{\mathcal{L}}
\newcommand{\difficulty}{\mathcal{D}}
\newcommand{\query}{\mathbf{q}}
\newcommand{\doc}{\mathbf{d}}
\newcommand{\docrel}{\mathbf{d^+}}
\newcommand{\docnrel}{\mathbf{d^-}}
\newcommand{\greyrule}{\arrayrulecolor{black!30}\midrule\arrayrulecolor{black}}
\newcommand{\sgreyrule}{\arrayrulecolor{black!30}\cmidrule{2-5}\arrayrulecolor{black}}

\newcommand{\fm}[1]{#1}

\newcommand{\nic}[1]{#1}
\newcommand{\edit}[1]{#1}
\newcommand{\sm}[1]{#1}

\AtBeginDocument{%
  \providecommand\BibTeX{{%
    \normalfont B\kern-0.5em{\scshape i\kern-0.25em b}\kern-0.8em\TeX}}}

\copyrightyear{2020} 
\acmYear{2020} 
\setcopyright{acmcopyright}\acmConference[SIGIR '20]{Proceedings of the 43rd International ACM SIGIR Conference on Research and Development in Information Retrieval}{July 25--30, 2020}{Virtual Event, China}
\acmBooktitle{Proceedings of the 43rd International ACM SIGIR Conference on Research and Development in Information Retrieval (SIGIR '20), July 25--30, 2020, Virtual Event, China}
\acmPrice{15.00}
\acmDOI{10.1145/3397271.3401094}
\acmISBN{978-1-4503-8016-4/20/07}

\begin{document}

\title{Training Curricula for Open Domain Answer Re-Ranking}

\author{Sean MacAvaney}
\affiliation{\institution{IR Lab, Georgetown University, USA}}
\email{sean@ir.cs.georgetown.edu}

\author{Franco Maria Nardini}
\affiliation{\institution{ISTI-CNR, Pisa, Italy}}
\email{francomaria.nardini@isti.cnr.it}

\author{Raffaele Perego}
\affiliation{\institution{ISTI-CNR, Pisa, Italy}}
\email{raffaele.perego@isti.cnr.it}

\author{Nicola Tonellotto}
\affiliation{\institution{University of Pisa, Italy}}
\email{nicola.tonellotto@unipi.it}

\author{Nazli Goharian}
\affiliation{\institution{IR Lab, Georgetown University, USA}}
\email{nazli@ir.cs.georgetown.edu}

\author{Ophir Frieder}
\affiliation{\institution{IR Lab, Georgetown University, USA}}
\email{ophir@ir.cs.georgetown.edu}

\fancyhead{}

\begin{abstract}
In precision-oriented tasks like answer ranking, it is more important to rank many relevant answers highly than to retrieve \textit{all} relevant answers. It follows that a good ranking strategy would be to learn how to identify the easiest correct answers first (i.e., assign a high ranking score to answers that have characteristics that usually indicate relevance, and a low ranking score to those with characteristics that do not), before incorporating more complex logic to handle difficult cases (e.g., semantic matching or reasoning). In this work, we apply this idea to the training of neural answer rankers using curriculum learning. We propose several heuristics to estimate the difficulty of a given training sample. We show that the proposed heuristics can be used to build a training curriculum that down-weights difficult samples early in the training process. As the training process progresses, our approach gradually shifts to weighting all samples equally, regardless of difficulty. We present a comprehensive evaluation of our proposed idea on three answer ranking datasets. Results show that our approach leads to superior performance of two leading neural ranking architectures, namely BERT and ConvKNRM, using both pointwise and pairwise losses. When applied to a BERT-based ranker, our method yields up to a 4\% improvement in MRR and a 9\% improvement in P@1 (compared to the model trained without a curriculum). This results in models that can achieve comparable performance to more expensive state-of-the-art techniques.
\end{abstract}

\maketitle

\section{Introduction}
\label{sec:intro}

Deep learning techniques are of recent interest to solve information retrieval tasks such as answer ranking~\cite{Mitra2017NeuralMF}. Most of such work focuses on designing neural network architectures that are effective at predicting answer relevance to a particular question, while comparatively little attention aims to find optimal training configurations for these networks. More so, existing literature often falls short of expressing the most basic settings of the training environment, such as the choice of the loss function and training sample selection procedures, two critical components needed for successful reproduction of results. In contrast, we focus on the neural rankers training process for information retrieval. In particular, we demonstrate that weighting training examples early in the learning process can yield significant benefits in the effectiveness of the neural ranker.

\begin{figure}[t!]
\centering
\vspace{1em}
\includegraphics[scale=2.7]{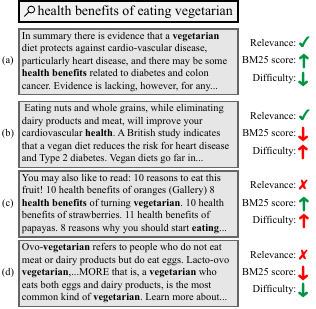}
\caption{Example of curriculum approach from MS-MARCO dataset (question ID 199776). In this example, we predict (a) is `easy' because it is relevant and has a high BM25 score. (d) is likewise `easy' weight because it is non-relevant and has a low score. (b) is a `difficult' sample because is relevant, yet has a low score due to the few term matches. We also predict (c) to be `difficult' because it is non-relevant, yet it has a high score. Our approach begins by weighting `easy' training samples high and `difficult' training samples low.}
\label{fig:overview}
\end{figure}

We motivate our approach with the simple intuition that some answers are easier to assess the relevance of than others. For instance, consider a question about the health impacts of vegetarianism (see Figure~\ref{fig:overview}). A passage written explicitly about this topic (e.g., (a)) should be relatively straightforward to identify, as it uses many of the terms in the question. This likely yields a high ranking position using conventional probabilistic approaches, such as BM25. A passage written about the health benefits of \textit{veganism} (a more strict version of vegetarianism) may also answer the question (b). However, it involves more complicated reasoning and inference (such as the understanding of the relationship between the two diets) and semantic understanding of the way in which the content is presented. Similarly, consider two non-relevant answers: one that matches most of the query terms (c) and one that does not (d). We argue that the former is more difficult for the ranking model to identify as non-relevant due to the large number of matching terms, and the latter is easier due to critical missing terms (e.g., health benefits).

While an ideal ranker would rank both (a) and (b) high, doing so we may add noise and complexity to the model that reduces the overall quality of ranking. Specifically, ranking (b) high may make it more difficult to identify (c) and (d) as non-relevant. Our method attempts to overcome this issue by forcing the ranker to focus primarily on the ``easy'' training samples before gradually moving on to learning to rank all training samples via training sample weighting.

We formulate this idea using the \textit{curriculum learning} (CL) framework~\cite{BengioEtAl2009}. Learning through a curriculum is an idea borrowed from cognitive sciences according to which the learning process follows a multi-step training path. Initially, the learning algorithm is trained by using simple examples and smooth loss functions. Then it is progressively fine-tuned so as to deal with examples and loss functions of increasing complexity. We instantiate the CL framework in the learning to rank domain by assigning different weights to training samples through a heuristic. In early stages of training, high weights are assigned to \textit{easy} training pairs while \textit{difficult} samples are given low weights. As training progresses, we gradually smooth this imbalance. Eventually, all training samples are weighted equally, regardless of the estimated difficulty.

To estimate the difficulty of question-answer pairs and to choose the training weight accordingly, we consider both information from an unsupervised baseline ranker (e.g., BM25) and the human-assessed relevance of the answer to the given question (see Figure~\ref{fig:overview}). When an unsupervised ranker is able to identify the example effectively (i.e., it ranks a relevant document high or a non-relevant document low) the training sample is considered as ``easy''. On the other hand, when the unsupervised ranker fails to correctly score them, the sample is considered as ``difficult''.

We show that our approach can be easily integrated into a neural ranking pipeline. We validate our approach using three weighting heuristics based on an unsupervised ranker using two leading neural ranking methods (BERT~\cite{devlin-19} and ConvKNRM~\cite{dai-18}). Our code is available for reproducibility.\footnote{\sm{\url{https://github.com/Georgetown-IR-Lab/curricula-neural-ir}}} Our results show significant ranking improvements when tested on three open-domain \edit{(i.e., not domain-specific)} answer ranking benchmarks: TREC Deep Learning (DL), TREC Complex Answer Retrieval (CAR), and ANTIQUE. These datasets vary in scale (hundreds of thousands of answers to tens of millions) and source of relevance information (graded or positive-only, human-annotated or inferred from document structure). We test using both pointwise and pairwise losses. In summary, our contributions are:

\begin{itemize}[leftmargin=*]
\item We propose a curriculum learning scheme for open-domain answer re-ranking.
\item We propose and evaluate three heuristics for weighting training samples while learning neural rankers, which utilize the ranking and score of the first-stage ranker.
\item We provide a comprehensive analysis of our proposed approaches for curriculum learning of neural rankers. Our results show the superiority of our approach as compared to standard weighting of training samples.
\item \edit{We show \fm{that} our proposed curricula are effective on three answer re-ranking datasets. On TREC DL, our approach yields up to a 3.6\% improvement in MRR and a 9.3\% improvement in P@1 for a BERT-based ranker. For TREC CAR, the curricula yield a 4.2\% and 3.7\% boost to R-Precision and MAP, respectively, and achieves comparable performance to a larger version of BERT. For ANTIQUE, our approach yields a 3.4\% and 6.0\% improvement in terms of MRR and P@1.}

\end{itemize}

\section{Background \& Related Work}
\label{sec:related}

\edit{In this section, we provide background information about neural ranking approaches (Section~\ref{sec:related:nir}) and prior work on curriculum learning (Section~\ref{sec:related:cl}).}

\subsection{Neural ranking}\label{sec:related:nir}
An ad-hoc neural ranking model maps a given query-document pair to a real-valued relevance score given the model parameters. For answer ranking, the question is treated as the query and answers are treated as the document. Through a set of training data, the parameters are optimized to maximize ranking performance on unseen data. Many neural architectures exist. They can be broadly classified as either \textit{representation-focused} or \textit{interaction-focused}~\cite{guo-16}.

Representation-focused models (also referred to as semantic matching models) learn mapping functions for the query and the document to a dense representation, and then compare these representations using a relatively simple comparison function (e.g., cosine similarity). Examples include DSSM~\cite{DSSM}, C-DSSM~\cite{CDSSM}, and ARC-I~\cite{ARCI}. These models rely on massive amounts of data (e.g., behavioral information from a query log) to learn semantic similarities. We do not focus on representation-focused models because of their general reliance on proprietary query logs, and their under-performance on standard test collections.

On the other hand, interaction-focused models (also referred to as relevance matching models) explicitly model patterns of query term occurrences in the document. DRMM~\cite{guo-16} models this concept by building a query-document similarity matrix where each cell represents the cosine similarity score between embeddings of each query term and document term. This allows the model to capture soft semantic term similarity (i.e., semantically-similar terms have a high score, and semantically-dissimilar terms have a low score). DRMM then models the term occurrences by building a histogram based on the similarity score for each query term and by producing a relevance score by feeding these histograms into a multi-layer perceptron. KNRM~\cite{xiong-17} works similarly, but replaces the hard histogram buckets with soft Gaussian-kernel-based buckets. Other approaches model term adjacency, such as MatchPyramid~\cite{pang-16}, DeepRank~\cite{pang-17}, PACRR~\cite{hui-18}, and ConvKNRM~\cite{dai-18}.

Contrary to recent critiques of the actual effectiveness of neural ranking architectures~\cite{Lin2018TheNH,Yang2019CriticallyET}, recent work with transformer-based contextualized language models (e.g., BERT~\cite{devlin-19}) on document and answer ranking have shown clear ranking superiority over prior baselines~\cite{macavaney:sigir2019-contextuallms,nogueira-19,Nogueira2019DocumentEB}. These methods exploit the distributional characteristics of a language learned through pre-training a model on tasks with more data available (e.g., masked language model and a next sentence prediction). Due to the self-attention mechanism of transformers, these models can also be considerd interaction-focused. \citet{macavaney:sigir2019-contextuallms} further demonstrated that signals from contextualized language models can be incorporated into other interaction-focused neural ranking architectures, boosting ranking performance beyond both transformer-based rankers and the non-contextualized interaction models.

\subsection{Curriculum Learning}\label{sec:related:cl}
Curriculum Learning (CL) can be considered a particular case of \textit{Continuation Methods}, generally used when the target objective function is non-convex and its direct optimization may lead the training to converge to a poor local minimum~\cite{ColemanEtAl1996,BengioEtAl2009}. The basic idea to overcome this problem through a curriculum approach is to organize the learning process as a path where the easiest training samples are presented first and the complexity of the following ones is gradually increased. This strategy allows the learner to exploit previously seen concepts to ease the acquisition of subsequent more difficult ones. CL approaches are proved successful for training neural networks in different domains such as NLP~\cite{CollobertEtAl2011,HuEtAl2014}, language models (not used for ranking tasks)~\cite{BengioEtAl2009}, image representation~\cite{ChenGupta2015},  network representation~\cite{QuEtAl2018}. To our knowledge, the only attempt to explore how CL methods can be exploited in the document ranking domain is the one by Ferro \emph{et al.}~\cite{FerroEtAl2018i} where authors exploit the curriculum learning strategy in a gradient boosting algorithm that learns ranking models based on ensembles of decision trees. The results reported by Ferro \emph{et al.} show that a curriculum learning strategy gives only a limited boost to the ranking performance of an ensemble of decision trees. Similar to our approach, Fidelity-weighted learning~\cite{Dehghani2017FidelityWeightedL} applies weights to training samples for learning ranking models. However, this approach focuses on estimating the quality of weak labels (namely, treating BM25 scores as labels), rather than the difficulty of training samples with higher-quality labels (e.g., human-annotated labels).

\edit{\citet{Sachan2016EasyQF} propose curriculum learning approaches for question answering, but in a closed-domain setting. In open-domain question answering, there are several challenges encountered, including that there is a much larger collection of answers to draw from (millions of answers) and multiple correct answers to a given question. Thus, we tackle this problem from an IR-perspective, utilizing signals from ranking models. Recently, \citet{Penha2019CurriculumLS} propose approaches for using curriculum learning to rank conversational responses, yielding up to a 2\% improvement in ranking effectiveness. The curricula proposed are specific to the domain of conversational responses and are non-trivial to apply to other domains. In contrast, we propose simple heuristics based on initial retrieval ranks and scores, and \fm{we show} their effectiveness across multiple ranking models, loss functions, and answer ranking datasets in an open-domain setting.}

\begin{table}
\centering\small
\caption{Table of symbols.}
\begin{tabular}{cl} \toprule
Symbol & Definition \\ \midrule
$R_\theta$ & Neural ranking function with parameters $\theta$ \\
$\loss$ & Loss function \\
$\difficulty$ & Training sample difficulty function \\
$W$ & Training sample weight function \\
$\query$ & Query (i.e., question) \\
$\doc$ & Document (i.e., answer) \\
$\docrel$ & Relevant document \\
$\docnrel$ & Non-relevant document \\
$\mathbf{D}$ & Set of ranked documents \\
$s$ & Manual relevance assessment score \\
$T$ & Collection of training data \\
$t$ & Training sample from $T$ \\
$i$ & Training iteration (epoch) \\
$m$ & End of curriculum iteration (hyperparameter) \\
\bottomrule
\end{tabular}
\label{tab:symbols}
\end{table}

\section{Methodology}
\label{sec:methodology}

We present our approach for applying curriculum learning to the training of neural rankers. At a high level, our approach applies a heuristic to determine the difficulty of a particular training sample. This difficulty estimation is then used for weighting training samples. In early stages of training, samples that the heuristic predicts as easy are given a higher weight, while samples predicted to be difficult are given a lower weight. Gradually, our approach eases off this crutch (controlled by a new hyper-parameter). Eventually, all training samples are weighted equally, regardless of the estimated difficulty.

Our approach allows for fair comparisons to an unmodified training process because no changes are made to the selection of the training data itself; the effect is only on the weight of the sample during training. Furthermore, this allows for an easy integration of our approach into existing training pipelines; no changes to the data selection technique are required, and the heuristics rely on information readily available in most re-ranking settings.

Our approach degrades into the typical training process in two ways: either (1) a heuristic can be used that gives every sample a weight of $1$, or (2) the hyper-parameter that drives the degradation of the approach to equal weighting can be set to immediately use equal weights. 

\subsection{Notation and preliminaries}
\label{sec:notation}
A summary of the symbols used is given in Table~\ref{tab:symbols}. Let an ad-hoc neural ranking model be represented as $R_{\theta}(\query,\doc)\in \mathbb{R}$, which maps a given query-document pair $(\query,\doc)$ to a real-valued relevance score given the model parameters $\theta$. For simplicity, we refer to questions as queries and answers as documents. Through a set of training data points $t \in T$ and a loss function $\loss(t)$, the model parameters $\theta$ are optimized to maximize the ranking performance. The training data sample $t\in T$ depends on the type of loss employed. Two common techniques employed for training neural rankers rely on pointwise or pairwise loss. For pointwise loss, training data consists of triples $t_{point} = \langle \query, \doc, s \rangle$, where $\query$ is a query, $\doc$ is a document, and $s$ is its relevance score, e.g., the relevance score given to the query-document pair by a human assessor. The loss for this sample often uses squared error between the predicted score and the relevance score $s$:
\begin{equation}
\loss^{point}(\query,\doc,s) = \big(s - R_{\theta}(\query,\doc)\big)^2
\end{equation}

On the other hand, pairwise loss uses two document samples for the same query (one relevant and one non-relevant), and optimizes to assign a higher score to the relevant document than the non-relevant one. Training triples for pairwise loss are represented as $t_{pair} = \langle \query, \docrel, \docnrel \rangle$, where $\query$ is the query, $\docrel$ is the relevant document, and $\docnrel$ is the non-relevant document. One common pairwise loss function is the softmax cross-entropy loss:
\begin{equation}
\loss^{pair}(\query,\docrel,\docnrel) = \frac{\exp\big(R_{\theta}(\query,\docrel)\big)}{\exp\big(R_{\theta}(\query,\docrel)\big) + \exp\big(R_{\theta}(\query,\docnrel)\big)}
\end{equation}

\subsection{Curriculum framework for answer ranking}
Let a difficulty function $\difficulty: T \mapsto [0,1]$ define a weight $\difficulty(t)$ for the training sample $t \in T$.
Without loss of generality we now assume that a high value of $\difficulty(t)$, i.e., a value close to $1$, represents an easy sample, while a low value, i.e., a value close to $0$, represents a difficult sample. Note that the heuristic $\difficulty(t)$ necessarily depends on the type of loss function employed: for pointwise loss, it estimates the difficulty for assigning the relevance score $s$ to $\langle \query, \doc \rangle$, while, for pairwise loss, it estimates the difficulty of scoring the relevant document pair $\langle \query, \docrel \rangle$ above the non-relevant pair $\langle \query, \docnrel \rangle$.

In our CL framework, during the first learning iteration, training samples are weighted according only to the difficulty function. To ease into the difficult samples, we employ a hyper-parameter $m$, which represents the training iteration at which to start to give every training sample equal weights.\footnote{We explore the importance of eventually converging to equal weights in Section~\ref{sec:eoc}.} Between the start of training and the $m$th training iteration, we linearly degrade the importance of the difficulty heuristic. More formally, we define the iteration-informed training sample weight $W_{\difficulty}(t,i)$ given the training iteration $i$ ($0$-based) as:
\begin{equation}\label{eq:weight}
W_\difficulty(t,i) =
\begin{cases}
  \difficulty(t) + \frac{i}{m} \big(1 - \difficulty(t)\big) & i < m \\
  1 & i \geq m
\end{cases}
\end{equation}

We then define a new $\difficulty$-informed loss function by including the iteration-informed weight into the standard pointwise or pairwise loss function:
\begin{equation}
\loss_\difficulty(t,i) = W_\difficulty(t,i)\, \loss(t)
\end{equation}

\subsection{Difficulty heuristics}
In a re-ranking setting, a simple source of difficulty information can come from the initial ranking of the documents. Probability ranking models like BM25 rely on term frequency and inverse document frequency to score documents. These characteristics should generally be easy for models to learn because they can learn to identify term frequency information (either directly, as is done by models like DRMM and KNRM, or implicitly, as is done by models like BERT through self-attention) and inverse document frequency, e.g., by down-weighting the importance of frequent terms. We postulate that it is inherently more difficult to perform semantic matching needed for identifying documents that have lower initial ranking scores. These scores are also easy to obtain, as they are readily available in a re-ranking setting. Thus, we use unsupervised ranking scores as the basis for our curriculum learning heuristics.

\paragraph{Reciprocal rank heuristic}
We define $\difficulty_{recip}$ as a measure of difficulty from the reciprocal of the rank at which answers appear in a ranked list. We assume that an answer placed higher compared to the other retrieved answers is ``easier'' for the ranker to place in that position. A high rank makes relevant documents easier and non-relevant documents harder. In the pointwise setting, relevant documents with a high reciprocal rank are considered ``easier'' than relevant documents with a low reciprocal rank because the unsupervised ranker assigned a higher score. Conversely, non-relevant documents with a high rank are considered ``harder'' than samples that are assigned a low rank. Given $\mathbf{d}$ from a set of ranked documents $\mathbf{D}$ for query $\query$ we have:
\begin{equation}
recip_{\query,\mathbf{D}}(\doc) = \frac{1}{rank_{\query,\mathbf{D}}(\doc)}
\end{equation}
With these conditions in mind, we define $\difficulty_{recip}$ for pointwise loss as:
\begin{equation}
\difficulty^{point}_{recip}(\query,\doc,s) =
\begin{cases}
  recip_{\query,\mathbf{D}}(\doc) & s > 0 \hspace{1em}\triangleright\textit{relevant} \\
  1 - recip_{\query,\mathbf{D}}(\doc) & s \leq 0 \hspace{1em}\triangleright\textit{non-relevant}
\end{cases}
\end{equation}
For pairwise loss, we define pairs that have a large difference between the reciprocal ranks to be very difficult (when the non-relevant document is higher) or very easy (when the relevant document is higher). When the reciprocal ranks are similar, we define the difficulty as moderate, with a difficulty close to $0.5$. This is accomplished by taking the difference between the scores and scaling the result within the range $[0,1]$:
\begin{equation}
\difficulty^{pair}_{recip}(\query,\docrel,\docnrel) = \frac
{recip_{\query,\mathbf{D}}(\docrel) - recip_{\query,\mathbf{D}}(\docnrel) + 1}
{2}
\end{equation}

\paragraph{Normalized score heuristic} An alternative to using the ranks of documents by an unsupervised ranker is to use the scores from these rankers. We define $\difficulty_{norm}$ as a measure of difficulty that uses the ranking score information. This allows documents that receive similar (or identical) scores to be considered similarly (or identically) in terms of difficulty. In the case of identical scores, $\difficulty_{norm}$ allows to improve the reproducibility of the CL approach compared to curricula that rely on rank~\cite{lin-19}. To account for various ranges in which ranking scores can appear, we apply min-max normalization by query to fit all scores into the $[0,1]$ interval\sm{, eliminating per-query score characteristics}. The integration of the normalized score $norm_{\query,\textbf{D}}(\doc)$ into both pointwise and pairwise rankers are similar to that of the reciprocal rank curriculum:
\begin{equation}
\difficulty^{point}_{norm}(\query,\doc,s) =
\begin{cases}
norm_{\query,\textbf{D}}(\doc) & s > 0 \hspace{1em}\triangleright\textit{relevant} \\
1 - norm_{\query,\textbf{D}}(\doc) & s \leq 0 \hspace{1em}\triangleright\textit{non-relevant}
\end{cases}
\end{equation}
\begin{equation}
\difficulty^{pair}_{norm}(\query,\docrel,\docnrel) = \frac
{norm_{\query,\textbf{D}}(\docrel) - norm_{\query,\textbf{D}}(\docnrel) + 1}
{2}
\end{equation}

\begin{figure}
\centering
\includegraphics[scale=0.4]{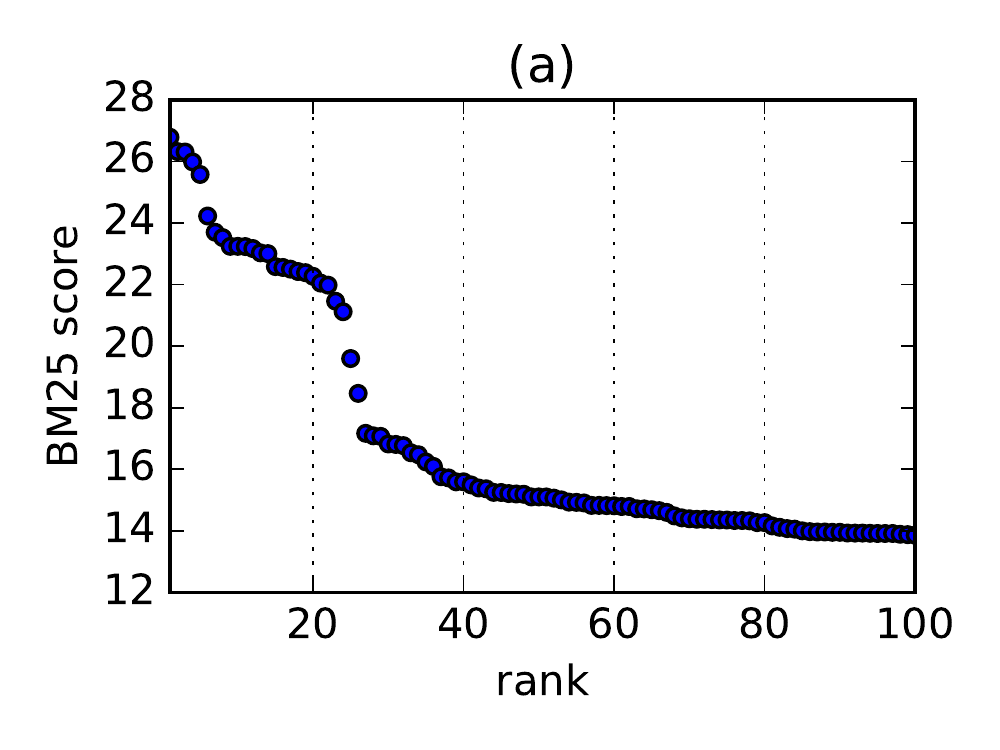}
\includegraphics[scale=0.4]{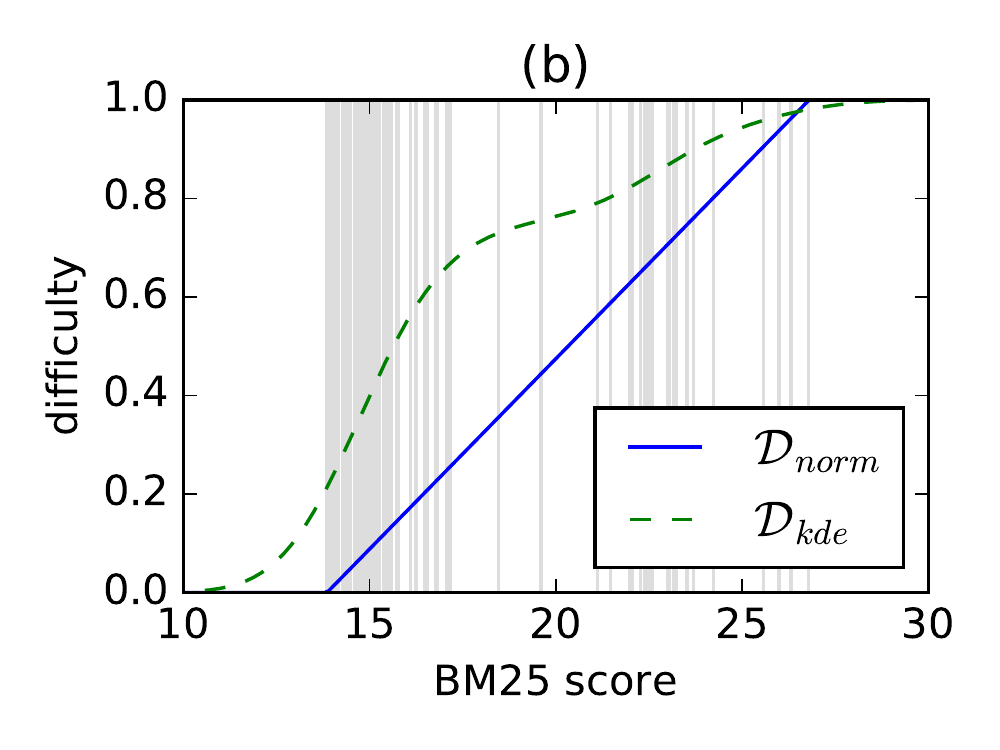}
\caption{(a) Example of BM25 scores exhibiting non-linear behavior; there are several answers with a much higher score than others and a long tail of lower-scored answers. (b)~Comparison between normalized score (solid blue) and KDE (dashed green) heuristic values by BM25 score. The grey vertical lines indicate the values from the initial ranking (from (a)). Scores are from MS-MARCO query 1000009 retrieved using Anserini's~\cite{Yang2018AnseriniRR} BM25 implementation.}
\label{fig:kde}
\end{figure}

\paragraph{Kernel Density Estimation (KDE) heuristic} The normalized score heuristic provides weighting based on ranking score, but it fails to acknowledge an important characteristic of ranking score values: they are often non-linear. For example, it is common for a handful of scores to be comparatively very high, with a long tail of lower scored answers (e.g., with fewer query term matches, see Figure~\ref{fig:kde}(a)). We hypothesize that it may be valuable to provide a greater degree of value distinction between scores in areas of high score density (e.g., in the long tail, around a score of 16 and below in Figure~\ref{fig:kde}(a)) and areas with relatively low score density (e.g., around a score of 20). To this end, we construct a Gaussian Kernel Density Estimation (KDE), with the bandwidth selected using Scott's Rule~\cite{scott2015multivariate}. We then define $\difficulty_{kde}$ by using the CDF of the kernel as the basis of difficulty measure:

\begin{equation}
\difficulty^{point}_{kde}(\query,\doc,s)=
\begin{cases}
KDE_{\query,\textbf{D}}(\doc) & s > 0 \hspace{1em}\triangleright\textit{relevant} \\
1 - KDE_{\query,\textbf{D}}(\doc) & s \leq 0 \hspace{1em}\triangleright\textit{non-relevant}
\end{cases}
\end{equation}
\begin{equation}
\difficulty^{pair}_{kde}(\query,\docrel,\docnrel) = \frac
{KDE_{\query,\textbf{D}}(\docrel) - KDE_{\query,\textbf{D}}(\docnrel) + 1}
{2}
\end{equation}
where $KDE_{\query,\textbf{D}}(\doc)$ yields the CDF score of the kernel density estimation for $\doc$. An example of the difference between $\difficulty_{norm}$ and $\difficulty_{kde}$ for a particular query is shown in Figure~\ref{fig:kde}(b). This approach has the added benefit of allowing a non-zero difficulty for positive samples that are not retrieved in the ranking list.

\vspace{0.3em}

\noindent\sm{In summary, we propose a curriculum learning approach for answer ranking. The approach weights training samples by predicted difficulty. We propose three heuristics for estimating training sample difficulty, based on the rank or score of an unsupervised ranking model.}

\begin{table*}
\centering
\caption{Dataset statistics. The values in parentheses indicate the average number of relevance judgments per query.}
\label{tab:datasets}
\begin{tabular}{lrrrrl}
\toprule
Dataset & \# Answers & Train Queries & Validation Queries & Test Queries & Test Judgments \\
&  & (judg. per query) & (judg. per query) & (judg. per query) \\
\midrule
TREC DL~\cite{Nguyen2016MSMA} & 8.8M & 504k (1.1) & 200 (0.7) & 43 (215.3) & Human (graded) \\
TREC CAR~\cite{Dietz2017TRECCA} & 30M & 616k (4.8) & 2.2k (5.5) & 2.4k (6.7) & Inferred (positive only) \\
ANTIQUE~\cite{Hashemi2019ANTIQUEAN} & 404k & 2.2k (11.3) & 200 (11.0) & 200 (33.0) & Human (graded) \\
\bottomrule
\end{tabular}
\end{table*}

\section{Experiments}
\label{sec:experiments}

\edit{
We conduct experiments on three large-scale answer ranking datasets --- namely TREC Deep Learning (DL)~\cite{Craswell2019OverviewTRECDL} (Section~\ref{sec:msmarco}), TREC Complex Answer Retrieval (CAR)~\cite{Dietz2017TRECCA} (Section~\ref{sec:car}), and ANTIQUE~\cite{Hashemi2019ANTIQUEAN} (Section~\ref{sec:antique}) --- and two neural rankers (Vanilla BERT~\cite{devlin-19,macavaney:sigir2019-contextuallms} and ConvKNRM~\cite{dai-18}) to answer the following research questions:
}

\begin{enumerate}
\item[RQ1] Are the proposed training curricula effective for training neural rankers for answer ranking? (Sections~\ref{sec:msmarco}--\ref{sec:antique})
\item[RQ2] Under which conditions is each curriculum more effective (e.g., amount and quality of training data, type of neural ranker trained, etc.)? (Sections~\ref{sec:msmarco}--\ref{sec:antique})
\item[RQ3] Is it important to shift to difficult samples, or can a ranker be successfully trained focusing only on easy samples? (Section~\ref{sec:eoc})
\item[RQ4] \edit{Is focusing on the easy samples first more beneficial to training than focusing on the hardest samples first? (Section~\ref{sec:anti})}
\end{enumerate}

Each dataset exhibits different characteristics (summarized in Table~\ref{tab:datasets}), as do the neural ranking architectures employed:
\begin{itemize}[leftmargin=*]
\item \textbf{``Vanilla'' BERT}~\cite{macavaney:sigir2019-contextuallms}. This model uses the sentence classification mechanism from a pretrained BERT contextualized model~\cite{devlin-19} (a deep transformer-based network) to model the semantic relevance between the question and answer. This model yields exceptional ranking performance at the expense of computational cost and is the foundation for most state-of-the-art answer ranking approaches. We test Vanilla BERT using both pointwise and pairwise loss, as defined in Section~\ref{sec:notation}. In line with~\cite{macavaney:sigir2019-contextuallms}, we initialize the model using \texttt{bert-base} (12-layer transformer pretrained on English text from~\cite{devlin-19}) and fine-tune using a learning rate of $2\times10^{-5}$ with the Adam optimizer.

\item \textbf{ConvKNRM}~\cite{dai-18}. This model learns the relationship between unigram and n-gram (via a convolutional neural network) similarity scores between the question and answer and combines the scores using Gaussian filters. This model yields competitive ranking performance and can be optimized for real-time ranking~\cite{Ji2019EfficientIN}. We use unigram to tri-gram encoding with cross-matching and 128 hidden nodes for score combination. Word vectors were initialized using 300-dimensional FastText~\cite{bojanowski2017enriching} word embeddings trained on WikiNews with subword information.
Based on preliminary experiments that showed that the ConvKNRM model fails to converge when trained using pointwise loss, we only test using pairwise loss.
We train the model using the Adam optimizer and a learning rate of $10^{-3}$. \fm{Furthermore, we} use the score additivity technique from~\cite{Yang2019CriticallyET}.
\end{itemize}

We train each model using training iterations consisting of 32 batches of 16 training samples uniformly selected over the re-ranking pool. We employ gradient accumulation when a training batch is unable to fit on a GPU (e.g., Vanilla BERT models). After each training iteration, the validation performance is assessed. We employ early stopping after 15 consecutive epochs with no improvement to the dataset-dependent validation metric. When training is early stopped, the model is rolled back to the version of that achieved a performance improvement. This yielded up to $130$ training iterations. We test our three proposed training curricula ($\difficulty_{recip}$, $\difficulty_{norm}$, and $\difficulty_{kde}$) on each of the datasets and neural rankers. We optimize the parameter $m$ i.e., end of curriculum learning epoch, by fine-tuning on the validation set. For each dataset, ranker, and loss combination, we test $m\in\{1,5,10,20,50,100\}$. \edit{To put performance of the neural rankers in context, we include the ranking effectiveness of Anserini's~\cite{Yang2018AnseriniRR} implementation of BM25 and SDM~\cite{Metzler2005AMR}, both with default parameters, tuned on the validation set (`Tuned'), and tuned on the test set (representing the optimal settings of parameters for this model, `Optimized').\footnote{Models tuned using a grid search: BM25 $k_1\in[0.1,4.0]$ by 0.1 and $b\in[0.0,1.0]$ by 0.05; SDM term, ordered and unordered weights $\in[0,1]$ by 0.1.} We also include relevant prior reported results and the optimal re-ranking of the results (i.e., sorting the original ranking list by relevance score, serving as an upper bound to re-ranking performance).}

\subsection{Web passage answer ranking}\label{sec:msmarco}

\begin{table}[b!]
\centering
\caption{Ranking performance on the TREC DL 2019 answer passage ranking task. \sm{Significant improvements in performance when using the training curricula (as compared to no curriculum) are indicated with $\uparrow$ (paired t-test $p<0.05$). There are no statistically-significant differences among the curricula.} The top result for each model are listed in bold.}
\label{tab:msmarco}
\begin{tabular}{cllrr}
\multicolumn{5}{c}{\bf TREC DL 2019} \\
\toprule
& Ranker & Training & MRR@10 & P@1 \\
\midrule


& \multirow{3}{*}{ConvKNRM} & Pairwise & 0.6159 & 0.4419 \\
&  & \phantom{- }w/ $\difficulty_{recip}$ & \bf 0.6834 & \bf 0.5581 \\
&  & \phantom{- }w/ $\difficulty_{norm}$ & 0.6514 & 0.5116 \\
&  & \phantom{- }w/ $\difficulty_{kde}$ & 0.6475 & 0.5116 \\
\greyrule
& \multirow{3}{*}{Vanilla BERT} & Pointwise & 0.8740 & 0.7907 \\
&  & \phantom{- }w/ $\difficulty_{recip}$ &\bf0.8942 &\bf0.8372 \\
&  & \phantom{- }w/ $\difficulty_{norm}$ & 0.8895 & 0.8140 \\
&  & \phantom{- }w/ $\difficulty_{kde}$ & 0.8857 & 0.8140 \\
\greyrule
& \multirow{3}{*}{Vanilla BERT} & Pairwise & 0.8477 & 0.7442 \\
&  & \phantom{- }w/ $\difficulty_{recip}$ & 0.8624 & 0.7674 \\
&  & \phantom{- }w/ $\difficulty_{norm}$ & 0.8581 & 0.7907 \\
&  & \phantom{- }w/ $\difficulty_{kde}$ &\bf0.8837 & $\uparrow$ \bf0.8372 \\

\midrule
\parbox[t]{2mm}{\multirow{10}{*}{\rotatebox[origin=c]{90}{Baselines}}}
& \multirow{3}{*}{BM25} & Default & 0.7024 & 0.5814 \\
& & Tuned & 0.6653 & 0.5349 \\
& & Optimized & 0.7555 & 0.6744 \\
\sgreyrule
& \multirow{3}{*}{SDM} & Default & 0.6276 & 0.4884 \\
& & Tuned & 0.6243 & 0.4884 \\
& & Optimized & 0.6667 & 0.5814 \\
\sgreyrule
& \multicolumn{2}{l}{\phantom{Top TREC Re-Ranking runs~\cite{Craswell2019OverviewTRECDL} }1.} & 0.907\phantom{0} & - \\
& \multicolumn{2}{l}{Top TREC Re-Ranking runs~\cite{Craswell2019OverviewTRECDL} 2.} & 0.882\phantom{0} & - \\
& \multicolumn{2}{l}{\phantom{Top TREC Re-Ranking runs~\cite{Craswell2019OverviewTRECDL} }3.} & 0.870\phantom{0} & - \\
\sgreyrule
& \multicolumn{2}{l}{Optimal Re-Ranker} & 0.9767 & 0.9767 \\

\bottomrule
\end{tabular}
\end{table}

\edit{We first demonstrate the effectiveness of our training curricula on the TREC Deep Learning (DL) 2019 answer passage ranking dataset, which uses the MS-MARCO collection and queries~\cite{Nguyen2016MSMA}. The training data for this dataset }consists of over a million questions collected from the Bing query log. A human annotator was presented a question and a list of 10 candidate answer passages. The annotator was asked to produce a written answer to these questions based on the passages and to indicate the passages that were most valuable in the production of this answer. For the purposes of passage ranking, these passages are considered relevant to the corresponding question. We note that this does not necessarily mean that all correct passages are annotated as relevant, nor it means that the \textit{best} passage is annotated (better answers could exist beyond the 10 shown to the annotator). \edit{To overcome this limitation, the TREC DL track manually judged the top retrieved passages for a subset of the test collection. This evaluation setting, which uses manual relevance judgments, is more suitable for evaluation than prior works that relied on incomplete relevance judgments (e.g.,~\cite{nogueira-19}).}
These incomplete training relevance labels also make this dataset suitable for our curriculum learning approach; answers ranked highly by an unsupervised ranker may be relevant, so down-weighting these samples during training may be beneficial.

\edit{We train our models using the official MS-MARCO list of training positive and negative relevance judgments. We use a held-out set of 200 queries for validation. We re-rank the official initial test set ranking,\footnote{\edit{Another evaluation setting for TREC DL is ``full ranking'', in which systems perform initial retrieval in addition to re-ranking. Since this work focuses on improving the effectivness of re-raning models rather than initial stage retrieval, we compare with other re-ranking submissions.}} and we use the official TREC DL manual relevance judgments for our evaluation and analysis.} Statistics about the training, development, and test sets are given in Table~\ref{tab:datasets}.

\sm{Since this work focuses on re-ranking, we evaluate using precision-oriented metrics, and leave recall to future work.} We use mean reciprocal rank at 10 (MRR@10) as the validation metric, as it is the official task evaluation metric. \edit{Although not included as an official task metric,} we also evaluate using P(recision)@1, which indicates the performance of the ranker in a realistic setting in which a single answer is given to a question.

We present the ranking performance for TREC DL in Table~\ref{tab:msmarco}. \edit{We observe that under all conditions, our proposed curricula out-perform the ranker when trained without a curriculum for both MRR and P@1 metrics. $\difficulty_{recip}$ outperforms the other curricula for ConvKNRM and pointwise Vanilla BERT, while $\difficulty_{kde}$ outperforms the other curricula for pairwise Vanilla BERT.}

\edit{When the model significantly under-performs well-tuned BM25 and SDM (ConvKNRM), we observe that the curricula can improve the ranking performance to approximately the level of these baselines. When the model is already doing substantially better (Vanilla BERT), our training curricula also yield a considerable boost to ranking effectiveness. The observation that our approach can improve the ranking effectiveness in both these cases is encouraging, and suggests that the approach is generally beneficial. When compared to the top TREC DL re-ranking results~\cite{Craswell2019OverviewTRECDL}, our approach performs favorably. Specifically, the top approach, \nic{namely pointwise Vanilla BERT with $\difficulty_{recip}$,} ranks second among the submissions. It is only narrowly exceeded by a much more expensive and complicated approach of pretraining a new BERT model from scratch using a different training objective. Our results indicate that this can be avoided by simply doing a better job weighting the training samples.}

\begin{figure}
\centering
\includegraphics[scale=0.55]{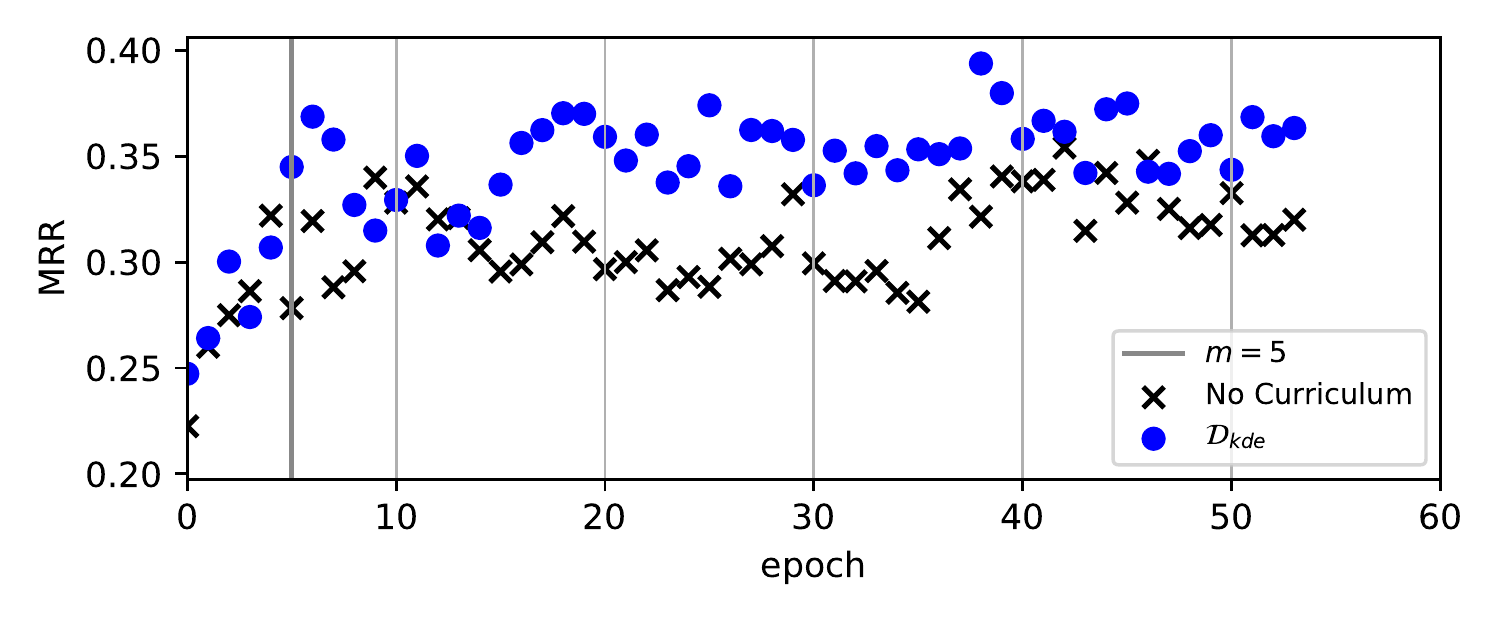}
\caption{Validation performance comparison between Vanilla BERT model trained with pointwise loss without a curriculum (black x) and with the $\difficulty_{kde}$ curriculum (blue circle) for TREC DL. The tuned $m$ parameter for the $\difficulty_{kde}$ curriculum used here is marked with a vertical line. While the variant without a curriculum quickly reaches optimal performance, the curriculum approach reaches a higher performance faster and offers a stronger foundation on which to continue training after the curriculum terminates.
}
\label{fig:validation_training}
\end{figure}

To gain a better understanding of how the curriculum benefits the training process, we compare the validation performance of the \nic{pointwise} Vanilla BERT model with \nic{the $\difficulty_{kde}$} training curriculum to the \nic{same} model when trained without a curriculum (Figure~\ref{fig:validation_training}). This reveals that when not using a curriculum, the validation performance peaks early, suggesting that it is overfitting to difficult examples. The curriculum, however, has even stronger early performance and is in a better position to incorporate difficult samples as training continues. Note that the tuned end of curriculum epoch is $m=5$ for this example, showing that the curriculum does not need to be in place for long to get these benefits. Also note that the training data were presented in the exact same order in both cases, showing the importance of weighting the loss effectively.

\subsection{Complex answer passage ranking}\label{sec:car}

\begin{table}
\centering
\caption{Ranking performance on the TREC CAR complex answer passage ranking task. Significant improvements in performance when using the training curricula (as compared to no curriculum) for each model are indicated with $\uparrow$ (paired t-test $p<0.05$, no significant reductions observed). \sm{For Pointwise loss, $\difficulty_{recip}$ significantly outperforms $\difficulty_{norm}$ in terms of MAP. There are no other significant differences among the training curricula.} The top results in each section are indicated in bold.}
\label{tab:car}
\begin{tabular}{cllrr}
\multicolumn{5}{c}{\bf TREC CAR} \\
\toprule
& Ranker & Training & R-Prec & MAP \\
\midrule

& \multirow{3}{*}{ConvKNRM} & Pairwise & 0.1081 & 0.1412 \\
&  & \phantom{- }w/ $\difficulty_{recip}$ & $\uparrow$ 0.1174 &  $\uparrow$ 0.1493  \\
&  & \phantom{- }w/ $\difficulty_{norm}$ & \bf $\uparrow$ 0.1258 & \bf $\uparrow$ 0.1572 \\
&  & \phantom{- }w/ $\difficulty_{kde}$ & $\uparrow$ 0.1227 & $\uparrow$ 0.1553 \\
\greyrule
& \multirow{3}{*}{Vanilla BERT} & Pointwise & 0.2026 & 0.2490 \\
&  & \phantom{- }w/ $\difficulty_{recip}$ &\bf $\uparrow$ 0.2446 &\bf $\uparrow$ 0.2864 \\
&  & \phantom{- }w/ $\difficulty_{norm}$ & $\uparrow$ 0.2370 & $\uparrow$ 0.2764 \\
&  & \phantom{- }w/ $\difficulty_{kde}$ & $\uparrow$ 0.2370 & $\uparrow$ 0.2795 \\
\greyrule
& \multirow{3}{*}{Vanilla BERT} & Pairwise & 0.2731 & 0.3207 \\
&  & \phantom{- }w/ $\difficulty_{recip}$ & $\uparrow$ 0.2914 & $\uparrow$ 0.3298 \\
&  & \phantom{- }w/ $\difficulty_{norm}$ &\bf$\uparrow$ 0.2921 &\bf$\uparrow$ 0.3307 \\
&  & \phantom{- }w/ $\difficulty_{kde}$ & $\uparrow$ 0.2844 & 0.3254 \\

\midrule
\parbox[t]{2mm}{\multirow{9}{*}{\rotatebox[origin=c]{90}{Baselines}}}
& \multirow{3}{*}{BM25} & Default  Settings & 0.1201 & 0.1563 \\
& & Tuned & 0.1223 & 0.1583 \\
& & Optimized & 0.1231 & 0.1588 \\
\sgreyrule
& \multirow{3}{*}{SDM} & Default  Settings & 0.1154 & 0.1463 \\
& & Tuned & 0.1099 & 0.1420 \\
& & Optimized & 0.1155 & 0.1459 \\
\sgreyrule
& BERT Large~\cite{nogueira-19} & & - & 0.335\phantom{0} \\
& BERT Base~\cite{nogueira-19} & & - & 0.310\phantom{0} \\
& PACRR~\cite{macavaney:irj2018-car} & & 0.146\phantom{0} & 0.176\phantom{0} \\
\sgreyrule
& \multicolumn{2}{l}{Optimal Re-Ranker} & 0.6694 & 0.6694 \\

\bottomrule
\end{tabular}
\end{table}

We also evaluate our curriculum learning framework on the TREC Complex Answer Retrieval (CAR) dataset~\cite{Dietz2017TRECCA}. \edit{To compare with prior work, we use version 1.0 of the dataset.} This dataset consists of topics in the form of a hierarchy of article headings (e.g., \textit{Green Sea Turtle \guillemotright{}  Ecology and behavior \guillemotright{} Diet}). A standard set of automatically-generated relevance judgments are provided by assuming paragraphs (passages) under a heading are relevant to the query corresponding to the heading. The automatic relevance assessments provide a large amount of training data, but can suffer from variable quality (e.g., some paragraphs are very difficult to match as they provide little context). This makes TREC CAR a good application of training curricula; it contains many positive relevance samples that are difficult to match. A set of manually-graded relevance assessments are also provided by TREC assessors. However, due to the shallow assessment pool used (due to the large number of topics), we opt to only evaluate our approach using the automatic judgments.\footnote{The track report suggests that the automatic judgments are a reasonable proxy for manual judgments as there is a strong correlation between the automatic and manual performance among the TREC submissions~\cite{Dietz2017TRECCA}.} \edit{We use TREC 2017 (\texttt{Y1}) training data with \texttt{hierarchical} relevance judgments. We also compare our results to the performance reported by~\cite{nogueira-19} and~\cite{macavaney:irj2018-car}, which use BERT and the PACRR neural ranking architecture augmented with entity embeddings for classification, respectively.}

Following previous work~\cite{nogueira-19}, we train and validate our models using the top $10$ results retrieved by BM25 and test on the top $1000$ results. We use the official task metric of R-Prec(ision) to validate our model. \edit{We also report MAP, another official metric for the task. We use these metrics rather than MRR and P@1 because CAR queries often need many relevant passages to answer the question, not just one.}

We present the performance of our training curricula on TREC CAR in Table~\ref{tab:car}. \edit{We observe that in all cases, the training curricula significantly improve the ranking effectiveness. When training rankers using pairwise loss, the $\difficulty_{norm}$ curriculum is most effective, and when training with pointwise loss, the $\difficulty_{recip}$ curriculum is most effective. In the case of ConvKNRM, without the curriculum, the ranker under-performs the unsupervised \nic{BM25 and SDM} baselines; with the curricula, it performs on-par with them. For Vanilla BERT, both when trained with pairwise and pointwise losses, the ranker outperforms the unsupervised baselines without the curricula, and improves significantly when using the curricula.}

\edit{When compared with the supervised baselines, \nic{i.e., BERT and PACRR}, the Vanilla BERT model trained with pairwise loss \nic{and $\difficulty_{norm}$ curriculum} ends up performing about as well as the large BERT baseline reported by~\cite{nogueira-19} \nic{(0.3307 versus 0.335 in terms of MAP, no statistically significant difference)}. This is a \fm{considerable} achievement because the Vanilla BERT model is half the size and about twice as fast to execute. This observation strengthens the case for using curricula when training because it can allow for similar gains as using a much larger model.}

\edit{The remaining gap between our trained models and the optimal re-ranker on the CAR dataset, however, indicates that there is still room for improvement in this task. In particular, a considerable challenge is ranking passages without much context highly without adding too much noise to the model.}

\subsection{Non-factoid question answering}\label{sec:antique}

\begin{table}
\centering
\caption{Ranking performance on the ANTIQUE non-factoid question answering task. Significant improvements in performance when using the training curricula (as compared to no curriculum) are indicated with $\uparrow$ (paired t-test $p<0.05$). \sm{There are no statistically-significant differences among the curricula.} The top results in each section are indicated in bold.}
\label{tab:antique}
\begin{tabular}{cllrr}
\multicolumn{5}{c}{\bf ANTIQUE} \\
\toprule
& Ranker & Training & MRR & P@1 \\
\midrule

& \multirow{3}{*}{ConvKNRM} & Pairwise & 0.4920 & 0.3650 \\
&  & \phantom{- }w/ $\difficulty_{recip}$ & \bf $\uparrow$ 0.5617 & \bf $\uparrow$ 0.4550 \\
&  & \phantom{- }w/ $\difficulty_{norm}$ & $\uparrow$ 0.5523 & $\uparrow$ 0.4450 \\
&  & \phantom{- }w/ $\difficulty_{kde}$ & $\uparrow$ 0.5563 & $\uparrow$ 0.4500 \\
\greyrule
& \multirow{3}{*}{Vanilla BERT} & Pointwise & 0.6694 & 0.5550 \\
&  & \phantom{- }w/ $\difficulty_{recip}$ & 0.6858 & 0.5850 \\
&  & \phantom{- }w/ $\difficulty_{norm}$ & 0.6888 & 0.5800 \\
&  & \phantom{- }w/ $\difficulty_{kde}$ & \bf 0.6953 & \bf 0.6000 \\
\greyrule
& \multirow{3}{*}{Vanilla BERT} & Pairwise & 0.6999 & 0.5850 \\
&  & \phantom{- }w/ $\difficulty_{recip}$ & \bf $\uparrow$ 0.7335 & \bf $\uparrow$ 0.6450 \\
&  & \phantom{- }w/ $\difficulty_{norm}$ & 0.7237 & 0.6250 \\
&  & \phantom{- }w/ $\difficulty_{kde}$ & 0.7244 & 0.6250 \\
\midrule
\parbox[t]{2mm}{\multirow{8}{*}{\rotatebox[origin=c]{90}{Baselines}}}
& \multirow{3}{*}{BM25} & Default Settings & 0.5464 & 0.4450 \\
& & Tuned & 0.5802 & 0.4550 \\
& & Optimized & 0.6035 & 0.4950 \\
\sgreyrule
& \multirow{3}{*}{SDM} & Default Settings & 0.5229 & 0.4050 \\
& & Tuned & 0.5377 & 0.4400 \\
& & Optimized & 0.5491 & 0.4700 \\
\sgreyrule
&\multicolumn{2}{l}{Best prior published (BERT)~\cite{Hashemi2019ANTIQUEAN}} & 0.7968 & 0.7092 \\
\sgreyrule
&\multicolumn{2}{l}{Optimal Re-Ranker} & 0.9400 & 0.9400 \\

\bottomrule
\end{tabular}
\end{table}

We also test our approach on the ANTIQUE non-factoid question answering dataset~\cite{Hashemi2019ANTIQUEAN}. \edit{Unlike TREC DL and CAR, ANTIQUE has more thoroughly annotated training queries, with an around 11 graded relevance judgments per query in the training and validation collections (crowdsourced) (see Table~\ref{tab:datasets}).} Furthermore, these include explicit labels for non-relevant answers, which are not present in the other two datasets. This more extensive annotation comes at the expense of scale, however, with far fewer queries to train upon. Nevertheless, ANTIQUE represents another valuable set of conditions under which to evaluate our curricula. \edit{We randomly sample from the top 100 BM25 results for additional negative samples during training.} We validate and test by re-ranking the top 100 BM25 results, and MRR as the validation metric and P@1 as a secondary metric. \edit{We use these two official task metrics (at relevance level of 3 or higher, as specified in~\cite{Hashemi2019ANTIQUEAN}) because the answers in ANTIQUE are self-contained, and these metrics emphasize correct answers that are ranked highly and first, respectively.}

We report the curricula performance on ANTIQUE in Table~\ref{tab:antique}. \edit{Similar to TREC DL, we observe that the $\difficulty_{recip}$ and $\difficulty_{kde}$ curricula are the most effective. For ConvKNRM, the curricula were able to overcome what would otherwise be a model that under-performs \nic{w.r.t. the BM25 and SDM} unsupervised baselines. For the \nic{pointwise and pairwise} Vanilla BERT models (which are already very effective), we observe gains beyond. In the case of pairwise-trained Vanilla BERT, the $\difficulty_{recip}$ curriculum significantly boosted ranking performance. Despite our efforts to reproduce the effectiveness of BERT reported in~\cite{Hashemi2019ANTIQUEAN}, we were unable to do so using the experimental settings described in that work. These results are still below that of an optional re-ranking, suggesting that there is still considerable room for improvement when ranking these non-factoid answers.}

\edit{To answer RQ1 (whether the training curricula are effective), we observed that for three answer ranking datasets (TREC DL, TREC CAR, and ANTIQUE) these curricula can improve the ranking effectiveness across multiple neural rankers and loss functions. We observe that when a ranker initially underperforms standard baselines (e.g., ConvKNRM), the performance is effectively boosted to the level of those baselines. When the ranker already exceeds these baselines (e.g., Vanilla BERT), we also observe a boost to ranking effectiveness, often comparable to or approaching the state-of-the-art while being considerably faster (e.g., using BERT Base instead of BERT Large) or less complicated (e.g., not requiring an expensive pre-training step). The observation that the curricula are effective in these various conditions suggests that these curricula are generally effective. To answer RQ2 (under what conditions each curriculum is effective), we observe that $\difficulty_{recip}$ and $\difficulty_{kde}$ are generally more effective for natural-language questions (TREC DL and ANTIQUE), while $\difficulty_{norm}$ is more effective for keyword/structured questions (TREC CAR). \sm{One possible} alternative explanation may be that the latter is better with weak relevance labels, as TREC CAR's relevance labels are obtained through a heuristic, rather than human annotators. It does not appear as if the amount of training data has an effect, as TREC DL and ANTIQUE exhibit similar characteristics, while having drastically different amounts of training data.}

\subsection{End of curriculum evaluation}\label{sec:eoc}

\begin{table}
\centering
\caption{Ranker performance when the curriculum always uses difficulty scores, and never assigns equal weight to all samples (i.e., $m=\infty$), and when employing the anti-curriculum ($\widehat\difficulty$). Significant reductions in performance are indicated with $\downarrow$ (paired t-test, $p<0.05$).}\vspace{-1em}
\label{tab:static}
{\renewcommand\arraystretch{1.29}
\begin{tabular}{lcrrr}


\multicolumn{5}{c}{\bf TREC DL} \\
\toprule
Ranker & Curriculum & $m$ & MRR@10 & P@1 \\
\midrule
\multirow{3}{*}{ConvKNRM} & $\difficulty^{pair}_{recip}$ & 20 & \bf 0.6834 & \bf 0.5581  \\
 & $\difficulty^{pair}_{recip}$ & $\infty$ &  0.6744 & 0.5581 \\
 & $\widehat\difficulty^{pair}_{recip}$ & 20 & $\downarrow$ 0.5414 & $\downarrow$ 0.3721 \\
\greyrule
\multirow{3}{*}{Vanilla BERT} & $\difficulty^{point}_{recip}$ & 10 & \bf 0.8942 & \bf 0.8372 \\
 & $\difficulty^{point}_{recip}$ & $\infty$ & 0.8205 & 0.7209 \\
 & $\widehat\difficulty^{point}_{recip}$ & 10 & 0.8527 & 0.7442 \\
\greyrule
\multirow{3}{*}{Vanilla BERT} & $\difficulty^{pair}_{kde}$ & 20 & \bf 0.8837 & \bf 0.8372 \\
 & $\difficulty^{pair}_{kde}$ & $\infty$ & $\downarrow$ 0.7752 & $\downarrow$ 0.6279 \\
 & $\widehat\difficulty^{pair}_{kde}$ & 20 & 0.8314 & 0.7209 \\
\bottomrule

\\
\multicolumn{5}{c}{\bf TREC CAR} \\
\toprule
Ranker & Curriculum & $m$ & R-Prec & MAP \\
\midrule
\multirow{3}{*}{ConvKNRM} & $\difficulty^{pair}_{norm}$ & 50 &\bf 0.1258 & 0.1572 \\
 & $\difficulty^{pair}_{norm}$ & $\infty$ & 0.1250 &\bf 0.1579 \\
 & $\widehat\difficulty^{pair}_{norm}$ & 50 & $\downarrow$ 0.1030 & $\downarrow$ 0.1324 \\
\greyrule
\multirow{3}{*}{Vanilla BERT} & $\difficulty^{point}_{recip}$ & 20 & 0.2446 & 0.2864 \\
 & $\difficulty^{point}_{recip}$ & $\infty$ & \bf 0.2475 & \bf 0.2894 \\
 & $\widehat\difficulty^{point}_{recip}$ & 20 & $\downarrow$ 0.2258 & $\downarrow$ 0.2709 \\
\greyrule
\multirow{3}{*}{Vanilla BERT} & $\difficulty^{pair}_{norm}$ & 10 & \bf 0.2921 & \bf 0.3307 \\
 & $\difficulty^{pair}_{norm}$ & $\infty$ & $\downarrow$ 0.2669 & $\downarrow$ 0.3103 \\
 & $\widehat\difficulty^{pair}_{norm}$ & 10 &  0.2837 & 0.3276 \\
\bottomrule

\\
\multicolumn{5}{c}{\bf ANTIQUE} \\
\toprule
Ranker & Curriculum & $m$ & MRR & P@1 \\
\midrule
\multirow{3}{*}{ConvKNRM} & $\difficulty^{pair}_{recip}$ & 100 & \bf 0.5617 & \bf 0.4550 \\
 & $\difficulty^{pair}_{recip}$ & $\infty$ & 0.5368 & 0.4100  \\
 & $\widehat\difficulty^{pair}_{recip}$ & 100 & 0.5366 & 0.4200 \\
\greyrule
\multirow{3}{*}{Vanilla BERT} & $\difficulty^{point}_{kde}$ & 10 &\bf 0.6953 &\bf 0.6000 \\
 & $\difficulty^{point}_{kde}$ & $\infty$ & $\downarrow$ 0.6139 & $\downarrow$ 0.4750 \\
 & $\widehat\difficulty^{point}_{kde}$ & 10 & 0.6677 & 0.5500 \\
\greyrule
\multirow{3}{*}{Vanilla BERT} & $\difficulty^{pair}_{recip}$ & 5 &\bf 0.7335 &\bf 0.6450 \\
 & $\difficulty^{pair}_{recip}$ & $\infty$ & 0.7158 & 0.6150 \\
 & $\widehat\difficulty^{pair}_{recip}$ & 5 & 0.7193 & 0.6200 \\
\bottomrule

\end{tabular}
}
\end{table}

We already observed in Figure~\ref{fig:validation_training} that when using a training curriculum, ranking performance not only peaks higher sooner, but also leaves the model in a better starting point for when all samples are weighted equally. However, an important question remains: Is it important to train with equal weight for all samples or can the difficulty weights be used exclusively? To this end, we perform a test that forgoes the curriculum convergence parameter $m$, directly using $\difficulty(\cdot)$ as the training sample weight, regardless of training iteration (i.e., $m=\infty$, or equivalently $W=\difficulty$ instead of Eq.~\ref{eq:weight}).

We report the performance for this experiment on each dataset for each top-performing curriculum in Table~\ref{tab:static} ($m=\infty$ setting). \edit{We observe that for all models on the TREC DL and ANTIQUE datasets, this approach leads to a drop in ranking effectiveness, suggesting that it is important to eventually perform equal sample weighting. Intuitively, this is important because if easy samples are always weighted higher than difficult samples, the model will be hindered in learning the more complicated function to rank difficult samples. Curiously, for TREC CAR, this setting sometimes leads to improved ranking effectiveness (though not a statistically significant improvement). \sm{One possible explanation is} that in situations where weak labels are used (rather than human-judged labels from top retrieved results), it may be better to always apply the weighting, as some inferred positive labels may be too distant from what the model will typically encounter at inference time.}

\edit{To answer RQ3 (whether shifting to difficult samples is important), we find that it is indeed beneficial to use our proposed weighting technique given in Eq.~\ref{eq:weight}, rather than always applying the difficulty weighting when using manually-assessed relevance labels.}

\subsection{Anti-curriculum: Hardest samples first}\label{sec:anti}

To test whether our intuitions that ``difficult'' samples are harmful during early phases of training, we conduct a study using an anti-curriculum, i.e., we train our models by weighting the more difficult samples higher than the easier samples. This was applied by swapping out the difficulty function $\difficulty$ with
$\widehat{\difficulty}(\cdot)=1-\difficulty(\cdot)$.
This has the effect of assigning high weights to samples that previously had low weights and vice versa. All usage of the difficulty function remains unchanged (e.g., the integration of the difficulty function into the weight function).

Table~\ref{tab:static} ($\widehat\difficulty$ setting) presents a ranking performance comparison when using the anti-curriculum. \edit{We observe that the anti-curriculum always reduces ranking effectiveness, sometimes significantly. In some cases, this can be rather severe; on TREC DL for Vanilla BERT (pairwise), the MRR is reduced by 0.0523 and P@1 is reduced by 0.1163, resulting in a model that underperforms one without any weighting at all. To answer RQ4, these results suggest that there is benefit to weighting the easiest samples higher first, rather than the more difficult samples.}

\section{Conclusions and Future Work}
\label{sec:conclusions}

We \fm{proposed} three weighting heuristics to train neural rankers using curriculum learning that boost performance when ranking answers on three datasets. Our proposed heuristics boost ranking performance through training samples weighting, without changing the sequence in which the data are presented to the model for training. \edit{
Generally, the reciprocal rank (RECIP) and kernel density estimation (KDE) curricula were the most effective, although when working with inferred relevance labels with TREC CAR, the normalized score (NORM) curriculum was more effective.
Although these gains were not always enough to achieve state-of-the-art performance, 
they were often able to approach the level of performance of larger or more complicated approaches (such as using BERT (large) or re-training BERT with a different pre-training objective). We experimentally showed that the convergence of the curriculum to equal weighting is important when manually-labeled test data are used, otherwise resulting in inferior effectiveness. Finally, we found that focusing on the easiest samples first (rather than the hardest samples) was also an important characteristic of this approach.}

\sm{Future work could explore alternative difficultly degradation functions or explore how well the method applies to other approaches, such as performing additional domain fine-tuning.} It could also combine the weighting strategies with more intelligent sampling approaches for relevant and non-relevant training pairs. We note that our proposed difficulty heuristics may be an effective starting point for sampling strategies. Even with more effective sampling, our weighting approach may be beneficial for ensuring that `easy' samples are ranked effectively. Another possible direction for future work could explore the use of self-paced learning techniques~\cite{Jiang2014SelfPacedLW,Jiang2015SelfPacedCL}, allowing the model to learn which training samples characteristics make a samples easy or difficult.

\section*{Acknowledgments}
Work partially supported by the ARCS Foundation. Work partially supported by the Italian Ministry of Education and Research (MIUR) in the framework of the CrossLab project (Departments of Excellence). Work partially supported by the BIGDATAGRAPES project funded by the EU Horizon 2020 research and innovation programme under grant agreement No. 780751, and by the OK-INSAID project funded by the Italian Ministry of Education and Research (MIUR) under grant agreement No. ARS01\_00917.

\bibliographystyle{ACM-Reference-Format}
\bibliography{biblio}

\end{document}